\documentclass[%
 reprint,
 amsmath,amssymb,
 aps,
prb,
]{revtex4-1}

\usepackage{graphicx}
\usepackage{dcolumn}
\usepackage{bm}

\begin{document}


\title{Twelve Inequivalent Dirac Cones in Two-Dimensional ZrB$_2$}

\author{Alejandro Lopez-Bezanilla$^{1}$}
\email[]{alejandrolb@gmail.com}
\affiliation{Theoretical Division, Los Alamos National Laboratory, Los Alamos, NM 87545, United States}
\affiliation{Materials Science Division, Argonne National Laboratory, 9700 S. Cass Avenue, Lemont, Illinois, 60439, United States}

\date{\today}
             
\begin{abstract}
Theoretical evidence of the existence of 12 inequivalent Dirac cones at the vicinity of the Fermi energy in monolayered ZrB$_2$ is presented. Two-dimensional ZrB$_2$ is a mechanically stable d- and p-orbital compound exhibiting a unique electronic structure with two Dirac cones out of high-symmetry points in the irreducible Brillouin zone with a small electron-pocket compensation. First-principles calculations demonstrate that while one of the cones is insensitive to lattice expansion, the second cone vanishes for small perturbation of the vertical Zr position. Internal symmetry breaking with external physical stimuli, along with the relativistic effect of SOC, is able to remove selectively the Dirac cones. A rational explanation in terms of d- and p-orbital mixing is provided to elucidate the origin of the infrequent amount of Dirac cones in a flat structure. The versatility of transition metal d-orbitals combined with the honeycomb lattice provided by the B atoms yields novel features never observed in a two-dimensional material.  

\end{abstract}

\maketitle

\section{Introduction}

In the search of the materials of tomorrow's technology, graphene was considered early on as a promising candidate for development and innovation\cite{Novoselov666}. The combination of an electronic band structure hosting the so-called massless Dirac fermions, together with a weak spin-orbit coupling (SOC), enable electrons to travel at relativistic speed over very long distances at room temperature. Governed by the Dirac equation, Dirac fermions emerge at isolated points in the irreducible Brillouin zone (iBZ), where an upper band touches a lower one. In undoped graphene, two Dirac cones are centered around the two non-equivalent K and K' points with a perfect electron-hole symmetry at the Fermi level. Each Dirac cone is therefore three-fold degenerate since they are shared by three contiguous BZ, namely a BZ contains the equivalent to two inequivalent cones.

Experimental results confirmed the ability of other two-dimensional (2D) group IV materials (silicene\cite{PhysRevLett.108.155501,2053-1583-3-1-012001}, germanene\cite{germanene14,0953-8984-27-44-443002}, stanene\cite{stanene15}) for hosting Dirac states, and theoretical predictions suggested new compounds with similar properties (graphynes\cite{PhysRevLett.108.086804}, 8-pmmn borophene\cite{PhysRevB.93.241405}). 
In most cases, 2D honeycombed lattices with Dirac cones are composed of a single type of element. Although binary compositions usually yield gapped electronic structures (BN\cite{doi:10.1021/nn301675f}, MoS$_2$\cite{PhysRevLett.105.136805}), branching away from the single-element material and search for multi-atom compounds may yield similar benefits.

More recently, the fabrication of a diverse array of 2D B monolayers has captivated a growing community of scientific experts focusing on their large list of exceptional physical properties\cite{Mannix1513,FengB}. B-based materials with additional electron-rich elements are potential candidates to expand the number of compounds with similar properties to the group IV materials. 
The discovery of high T$_c$ superconductivity in MgB$_2$ \cite{ISI:000167194300040} caused an immediate interest to find B-based superconductors. The viability of this compound, similar to other planar diborides\cite{doi:10.1021/ic50030a027} such as B$_2$O  and B$_2$Be, relies in that it is unsymmetrical isoelectronic to C, with four valence electrons per atom in average. 2D B-monolayers incorporating elements with larger principal quantum numbers may benefit of the richness of d orbitals. Thus, first-principles calculations predicted the existence of massless fermions in monolayered TiB$_2$ \cite{PhysRevB.90.161402}.  Polymers containing both boron and transition metals atoms have been prepared, proving a high degree of delocalization via boron and metal atoms\cite{doi:10.1021/om010258d}. 

\begin{figure}[htp]
 \centering
   \includegraphics[width=0.5 \textwidth]{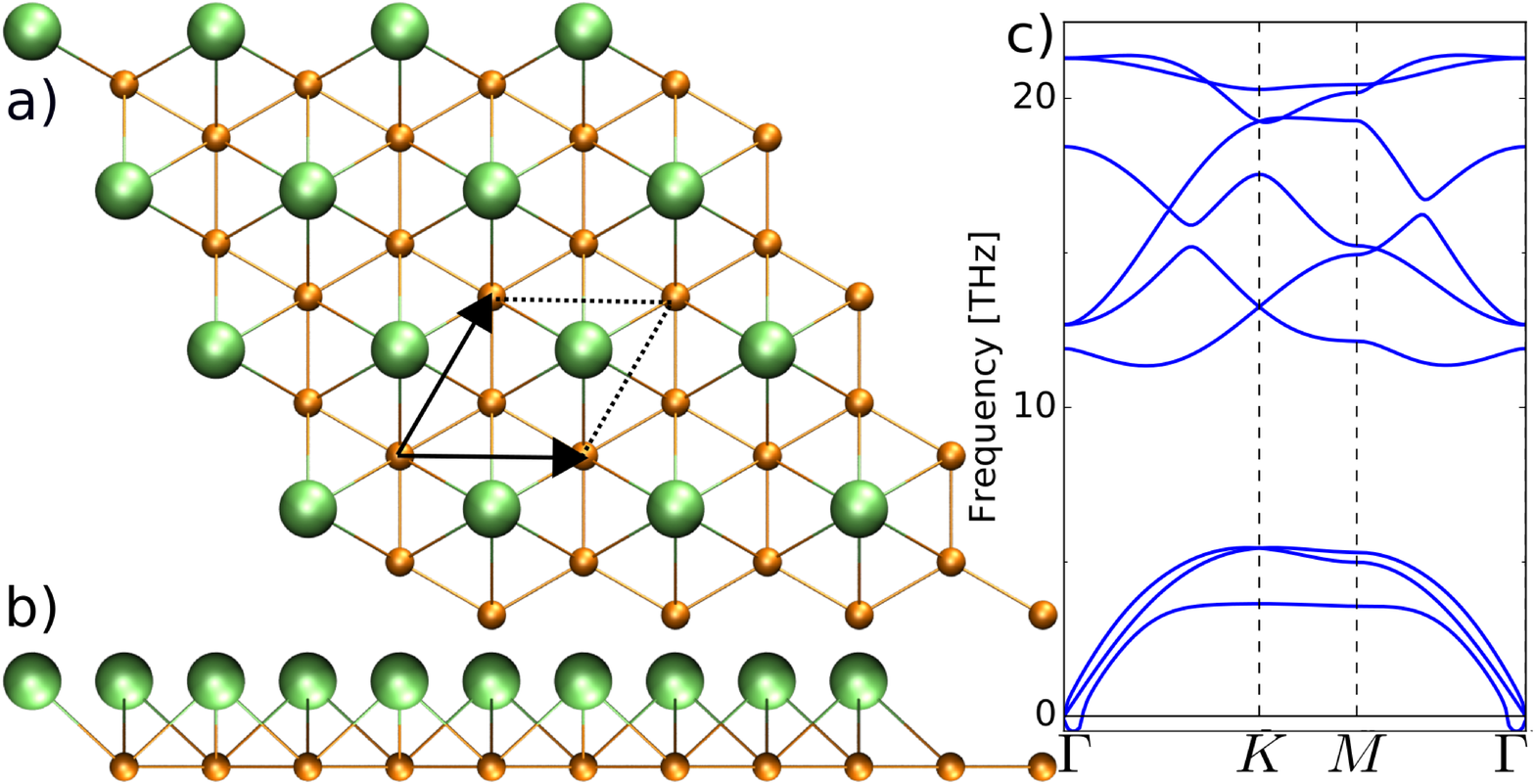}
 \caption{Top a) and side view of a ball-and-stick representation of the hexagonal cell of monolayered ZrB$_2$. Zr atoms sit on the center of the hexagons of the B atom honeycomb lattice . The hexagonal unit cell containing two atoms of B and one atom of Zr is shown in a). c) Phonon spectrum of monolayered ZrB$_2$}  
 \label{fig0}
\end{figure}

In this paper the potential scope of diborides is expanded by describing the electronic properties of metal diboride monolayered ZrB$_2$. Free-standing ZrB$_2$ is triangular borophene with Zr atoms in substitutional doping, and is predicted to exhibit 12 non-degenerate Dirac cones in the vicinity of Fermi level with a small compensated electron-hole pocket. This material features an infrequent property in a hexagonal lattice such as the displacement of the Dirac cones to lines joining two high-symmetry points of the iBZ. Both hexagonal $\beta$-graphyne\cite{PhysRevLett.108.086804} and TiB$_2$\cite{PhysRevB.90.161402} have been reported to exhibit six low symmetry Dirac points in the first BZ, and the use of Rashba SOC has been proposed\cite{PhysRevB.93.035401} for splitting them into 12. Instead of resorting to the spin degree of freedom, we propose the interplay of p and d orbitals to generate six groups of two inequivalent Dirac cones in the low energy spectrum of a P6mm space group material. Zr atom d-orbitals yield the formation of the Dirac cones with some contribution from the B p$_z$ orbitals. For the sake of clarity, it must be noted that the 12 cones refer to both the 12 upper and 12 lower cones that in pairs are created in the vicinity of the Fermi level.

Contrary to the current trend of identifying new materials with improved functionalities by means of high-throughput screening of thousands of compounds, here we present a one-at-a-time material study to provide a detailed analysis of a nanostructure designed upon examination of the physico-chemical properties of its constituents.

\section{Computational Methods}

Calculations were performed with the {\tt SIESTA} code. A double-$\zeta$ polarized basis set was used and the radial extension of the orbitals had a finite range with a kinetic energy cutoff of 50 meV. Electron exchange and correlation was described within the Perdew, Burke, and Ernzerhof \cite{PhysRevB.54.11169} scheme of the generalized gradient approximation. The integration in the k-space was performed using a 32$\times$32$\times$1 Monkhorst-Pack \cite{PhysRevB.13.5188} k-point mesh centered at $\Gamma$-point. To determine accurately the location of the Fermi level, a set of 320$\times$320$\times$1 k-points were used in a single-energy point calculation. Lattice constants and atomic coordinates were fully relaxed until the residual forces were smaller than 10$^{-3}$ eV/\AA. Phonon diagram dispersion was calculated with the {\tt PHONOPY} code \cite{phonopy} using the force-constant method, and the dynamical matrices were computed using the finite differences method in large supercells. Single-energy point SOC calculations were carried out with {\tt Elk} code\cite{Elk} with similar effective parameters.

\section{Results and Discussion}

In 2D ZrB$_2$, the honeycomb lattice of B atoms exhibits a B-B distance of 1.85\AA, and the Zr atoms sit over the hexagon hole at 1.48\AA\ (see Figure \ref{fig0}a and b). The lattice parameter is 3.17\AA. With its vacant p$_z$ orbital, three-fold coordinated B is an electron deficient atom and a strong p-electron acceptor.

\begin{figure}[htp]
 \centering
   \includegraphics[width=0.49 \textwidth]{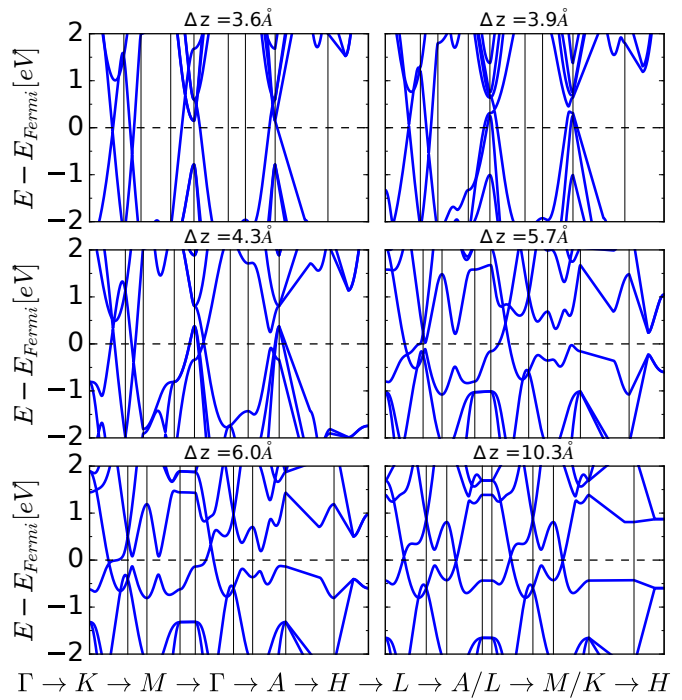}
 \caption{Evolution of the electronic states of ZrB$_2$ from bulk to monolayer for increasingly larger inter-layer distance $\Delta$z. The k-point path in the  Brillouin zone is as specified in the list of high symmetry points, each corresponding to vertical lines in the panels.}  
 \label{fig1}
\end{figure}

Before inspecting the electronic properties of the free-standing ZrB$_2$ monolayer, it is worth analyzing their origin from the bulk configuration. Vertical AA stacking of single hexagonal layers yields bulk ZrB$_2$, a ceramic material with a hexagonal covalent structure possessing good thermal and electrical conductivities. The crystal structure belongs to the space group P6/mmm, where the Zr atoms sit in between two hexagons of two B honeycombed network. Temperature dependent resistivity and high frequency susceptibility measurements in ZrB$_2$ revealed a superconducting transition at 5.5 K\cite{Gasparov2001}. Thin films of ZrB$_2$ can be grown by chemical vapour epitaxy on Si wafers, to additionally provide an electrically conductive substrate for GaN\cite{1347-4065-40-12A-L1280}, or the synthesis of monolayered materials \cite{2053-1583-4-2-021015}.

The series of electronic band diagrams in Figure \ref{fig1} shows the evolution of the electronic states of ZrB$_2$, from bulk to monolayer, with successive increases of the distance $\Delta z$ between stuck layers with no geometric optimization of the atomic positions. The origin of the good bulk conducting properties are clearly explained by the numerous metallic bands, including some band-crossing forming Dirac cones at the vicinity of the Fermi level, such in the $\Gamma-K$ line and $K-M$ lines. For a small increase of $\Delta z= 0.3$\AA, a first gap appears in the $H-L$ line at 0.4 eV, while in the vicinity of the $A$ point new lines cross the Fermi level and a Dirac cones is defined. This cone  lowers in energy at $\Delta z= 4.3$\AA, and backs up at $\Delta z= 5.7$\AA\ while the cone in the $K-M$ line vanishes. At a $\Delta z= 6.0$\AA\ interlayer separation only four metallic bands remain, and for $\Delta z= 10.3$\AA\ the Dirac cones defining the properties of monolayered ZrB$_2$ are finally defined.

\begin{figure}[htp] 
 \centering
  \includegraphics[width=0.49 \textwidth]{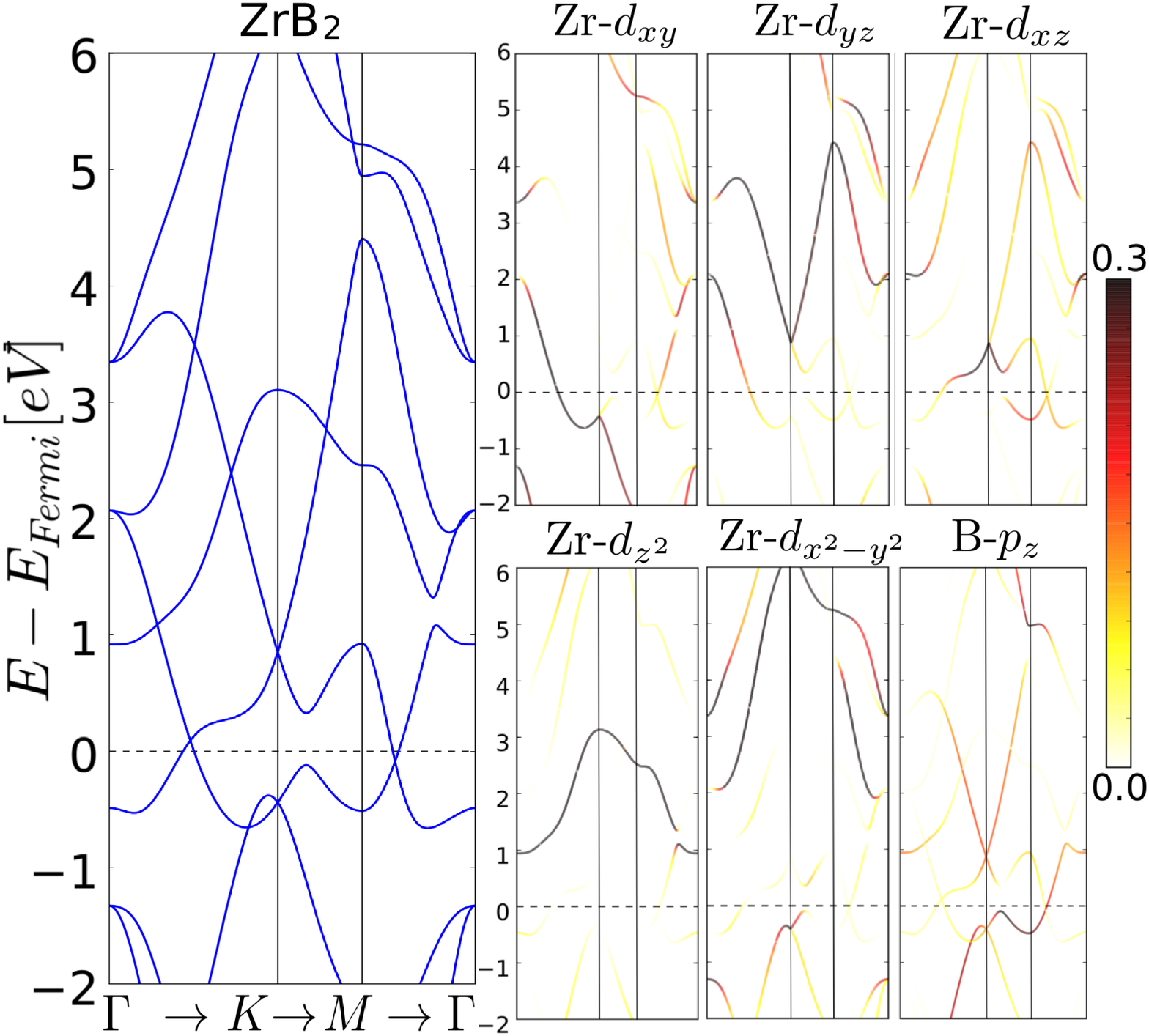}
 \caption{Left panel: Electronic band diagram of the irreducible Brillouin zone of free-standing monolayered ZrB$_2$. Right panels: color-resolved projected density of states onto each Zr d-orbitals and B p$_z$ orbital.}
 \label{fig2}
\end{figure}

Isolating a single layer reduces the symmetry to P6/mm, and the BZ of interest is the area delimited by the $\Gamma$, $M$, and $K$ points in the hexagonal reciprocal lattice. The monolayer is dynamically stable as verified by the absence of negative frequencies in the phonon spectrum shown in Figure \ref{fig0}c. Only one of the acoustic modes undergoes a tiny deviation towards negative frequencies close to the $\Gamma$ point.  
9 modes are extended over a frequency range of $\sim$21 THz. The in-plane acoustic branches are characterized by linear dispersions at low momentum near the center of the BZ. The out-of-plane phonon branches exhibit non-linear energy dispersions at the zone edge. While most acoustic modes exhibit $\sim$6 eV dispersive branches in the lower quarter of the spectrum, a frequency gap separates optical modes in the upper half, which are equally dispersive spanning altogether a range of $\sim$10 THz.

The band structure of the hexagonal 2D lattice is displayed in the left panel of Figure \ref{fig2}. Two bands form a Dirac cone in the $\Gamma-K$ line slightly above the Fermi level creating a small electron pocket. To compensate the lack of electrons, a similar hole pocket is formed upon two bands creating a second Dirac cone slightly below the Fermi energy level in the $M-\Gamma$ line. 

\begin{figure}[htp]
 \centering
   \includegraphics[width=0.47 \textwidth]{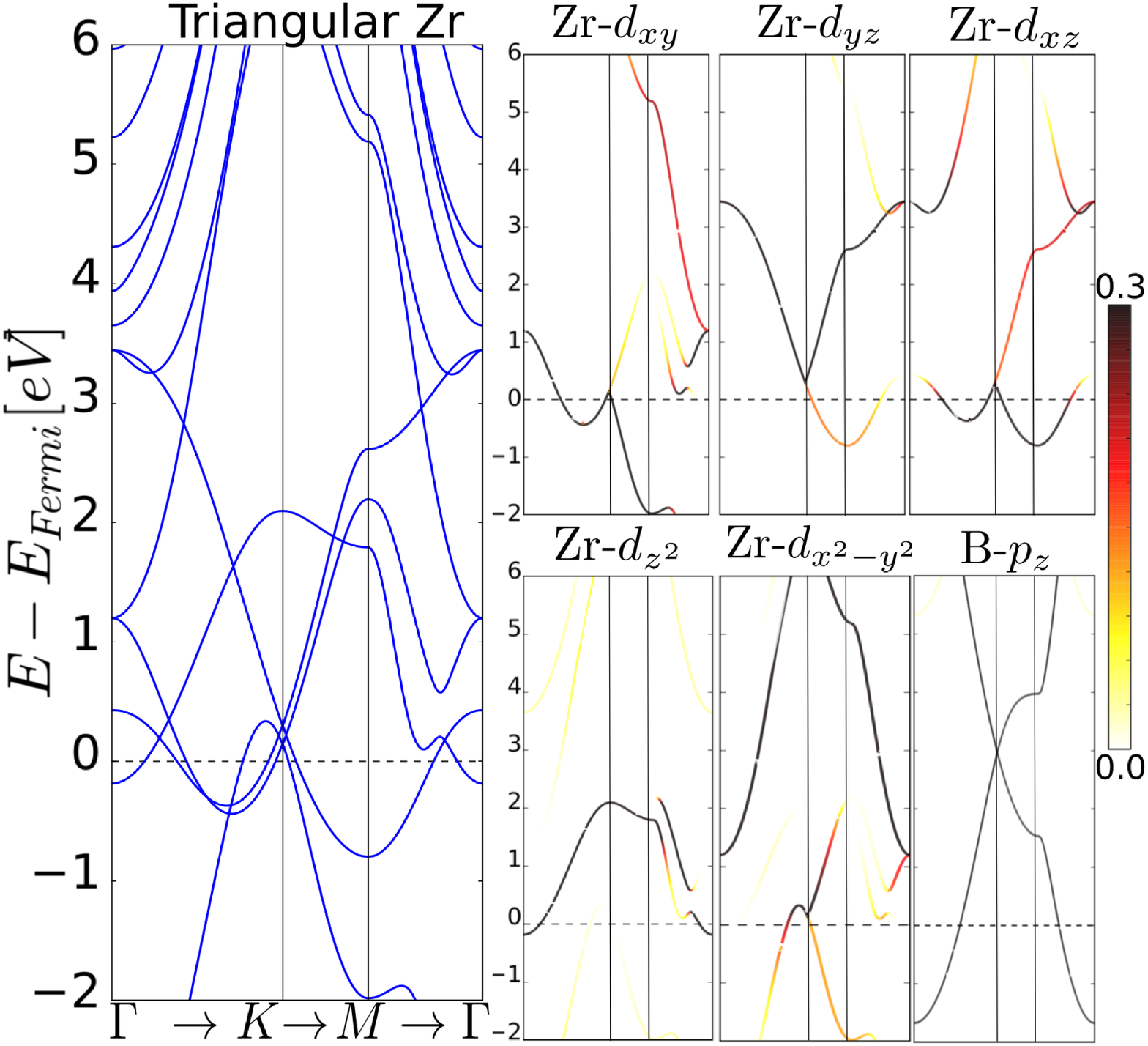}
 \caption{Left panel: Electronic band diagram of the triangular lattice composed of uniquely the Zr atoms of the ZrB$_2$ lattice at fixed positions. Color-resolved panels correspond to the projected density of states of the d-orbitals, and allow to analyze the contribution of the d-orbitals separately to each electronic state. Also the p$_z$ orbital of B atoms in an all-B honeycomb lattice is shown. }  
 \label{onlyZr}
\end{figure}

In the hexagonal pyramid configuration, monolayered ZrB$_2$ belong to the C$_{6v}$ point group. According to this symmetry representation, the five Zr d-orbitals are grouped in three groups, where two in-plane orbitals are degenerate (e$_2$), two out-of-plane orbitals also form a degenerate pair (e$_1$), and d$_{z^2}$ remains non-degenerate (a$_1$).
To corroborate this picture and elucidate the origin of each Dirac cone, we resort to color-resolved projected density of state (PDoS) diagrams, which allow to visualize the independent contribution of each atomic orbital to the electronic states. At $K$ point two band-crossing occur with a difference in energy of 1.29 eV. In the valence band, the d$_{xy}$ and the d$_{x^2-y^2}$ orbitals intersect in a two-fold degenerate point (e$_2$), while a Dirac cone at 0.87 eV in the conduction band is created upon the d$_{yz}$ and the d$_{xz}$ Zr orbitals meet in a degenerate pair (e$_1$). The d$_z^2$ orbital stands alone at 3.13 eV (a$_1$). Note that the intensity of the lines depends on the contribution but also on the normalization factor and, although some states may appear weakly represented, the presence with light colors at the Fermi level guarantees their contribution to the Dirac cones.

The PDoS show evidence that the $\Gamma-K$ Dirac cone at the Fermi energy is  created predominantly by the combination of Zr d-orbitals, whereas in the $M-\Gamma$ cone B p$_z$ orbitals have a greater contribution. The former derives from the d$_{xy}$, d$_{xz}$, and d$_{yz}$ orbitals, while the latter is mainly composed of d$_{xz}$ and d$_{yz}$ states hybridized with the B atom p$_z$ orbital. A reminiscence of the boron honeycomb lattice is evident in the contribution of the B p$_z$ orbitals to the Dirac cone at the $K$ point located at $\sim$0.9 eV, in whose formation the d$_{yz}$ for the upper cone and the d$_{xz}$ for the lower one also participate. The two d$_{z^2}$ and d$_{x^2-y^2}$ orbitals form anti-bonding states at high energy levels and do not contribute in any case to the formation of any Dirac cone. However, and as discussed below in more detail, although B atoms are crucial in the formation of the Dirac points, each cone presents different sensitivity to hybridization changes. 

\begin{figure}[htp]
 \centering
 \includegraphics[width=0.45 \textwidth]{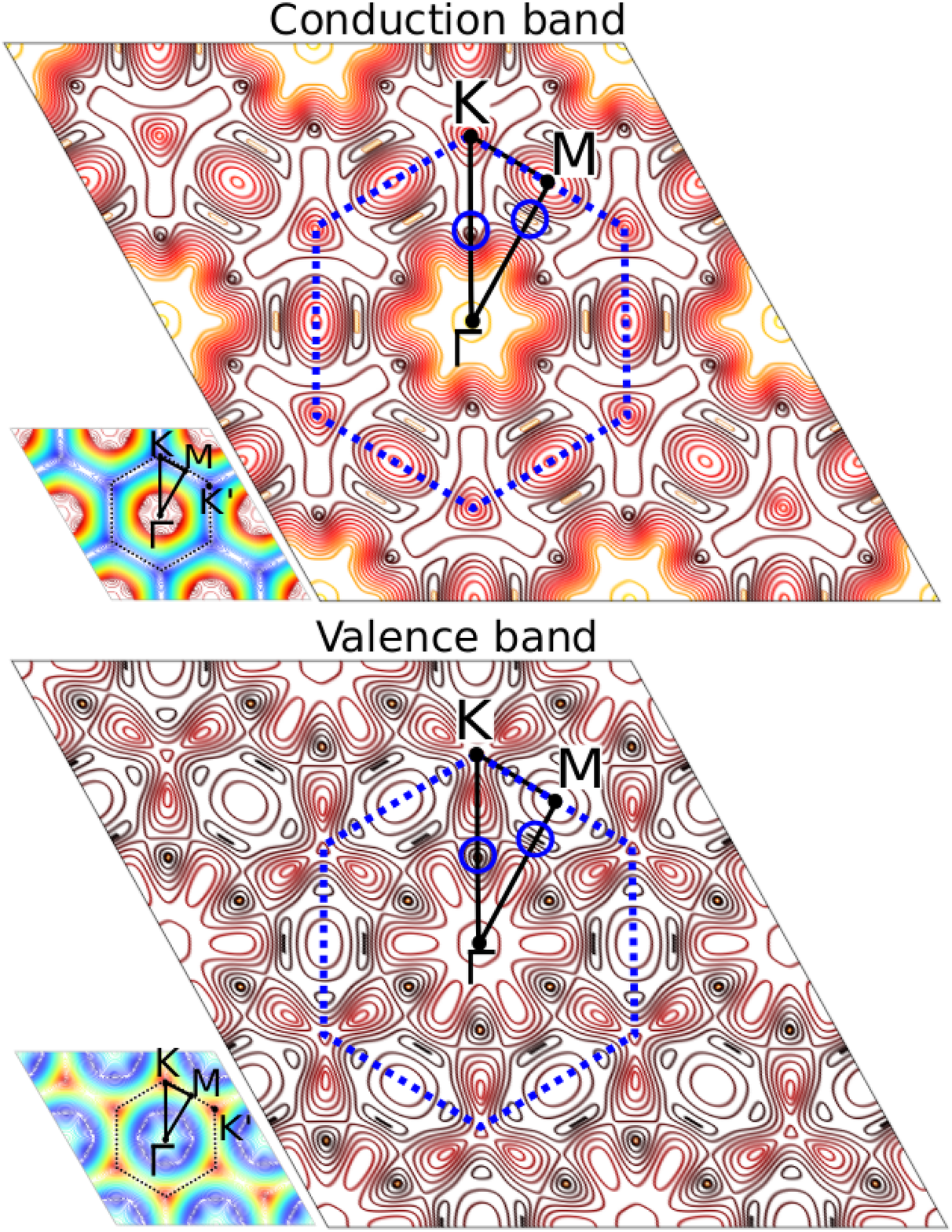}
 \caption{Constant-energy contour plots of the conduction and valence bands of ZrB$_2$ in the first (enclosed by dashed hexagon) and neighbouring Brillouin zones (BZ). Smaller plots correspond to the same type of diagram of graphene electronic bands. $\Gamma$, M, and K points delimit the irreducible BZ. Circles point out the location of the ZrB$_2$ Dirac points. By applying the C$_{6v}$ group symmetry operations, a total of 12 Dirac cones are obtained, in contrast to the three groups of two obtained in graphene at $K$ and $K'$ points.}
 \label{fig5}
\end{figure}

To further support this picture of d- and p-hybridized derived Dirac states, the band diagram of the triangular lattice formed uniquely by Zr atoms at fixed positions, and with all B atoms removed, is plotted in Figure \ref{onlyZr}. A single panel showing the band developed by the lateral superposition of B atom p$_z$ orbitals with no Zr atoms sitting on the hexagons is included. Interestingly, bands are shifted in energy and conserve the main features of the complete lattice, except the Dirac cones at the high-symmetry lines. The same group of four bands crossing at the $K$ point are distinguishable although within a much reduced energy range of 0.16 eV. In the $M-\Gamma$ line, the band intersection observed at 0.17 eV is composed of the d$_{xz}$ orbital for the running up band, and of the d$_{xy}$ and d$_{z^2}$ orbitals at the bending region of the running low band. Upon introducing the B atoms, p$_z$ orbitals mix with low-energy d orbitals in this region of the BZ and allow the formation of the Dirac cone. 
Similarly, in the first half of the $\Gamma-K$ line, there is no running up band that may yield the Dirac cone observed in the diborade lattice, other than the d$_{z^2}$ derived band which is known from the diboride diagram to not mix with any other state. Furthermore, the d$_{xz}$ state is developed from 0.43 eV down to -0.37 eV across that region, contrary to the observation in Figure \ref{fig2}, where it is developed from $\Gamma$ to $K$ with positive gradient. Such a change in the slope of the band can be ascribed to the presence of the B atom p$_z$ orbitals when they enter the formation of the diboride lattice, which also causes the d$_{z^2}$ and d$_{x^2-y^2}$ orbital derived bands to shift up in energy. 
This proves that the Zr-B orbitals interplay is crucial in the formation and strength of both Dirac cones, allowing the Zr-B hybridization to locate both Dirac points in the vicinity of the Fermi level, and the electronic delocalization over the complete unit cell. 

Applying the symmetry operations of the C$_{6v}$ point group to the iBZ, two dimensional ZrB$_2$ exhibits 12 inequivalent Dirac cones, which is a remarkable difference with respect to any previously described hexagonal material. This is displayed in the contour line plots of both the valence and conduction band shown in Figure \ref{fig5}, where the same bands of graphene have been included for comparison. Indeed, rotating the iBZ six times to complete the whole BZ of the hexagonal lattice (pointed out by the dashed hexagons), six groups of two Dirac points are reproduced. Note that applying the same operations on the graphene BZ yields three groups of a pair of K and K' points hosting the corresponding cones. 

\begin{figure}[htp]
 \centering
 \includegraphics[width=0.47 \textwidth]{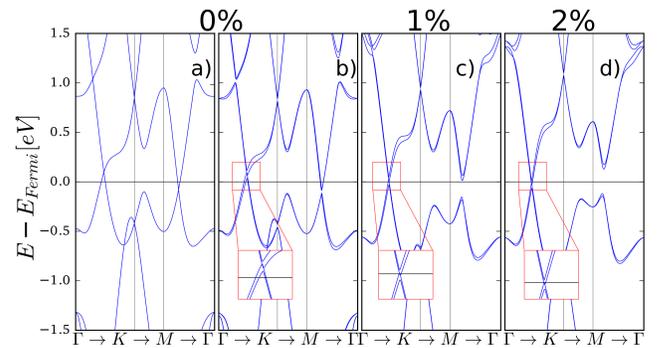}
 \caption{Non-relativistic band diagram in a) is shown for comparison with the spin-orbit coupling calculation in b). c) and d) shows the evolution of the monolayered ZrB$_2$ electronic states under tensile strain. Increments are shown in hexagonal lattice parameter percentage. Zr atom vertical position changes as the honeycomb lattice expands, and the Dirac point in the $M-\Gamma$ line vanishes. The superposition of electronic bands upon SOC-induced band splitting prevents from the creation of a gap in the $\Gamma \rightarrow K$ line. 
}
 \label{fig6}
\end{figure}

Due to the large size of the Zr atom, SOC is susceptible to introduce some changes in the electronic states. SOC was checked and the resulting band diagram is shown in Figure \ref{fig6}b. meV large band gaps are noticeable at all band crossing, while the rest of the electronic states barely differs from the non-relativistic band structure. It is worth noting that electron-hole pockets are preserved and the band splitting causes a distortion of the Dirac cone in the $\Gamma-K$ line that closes effectively an, in principle, small gap by overlapping the bands, as shown in detail in the zoom-in inset.

Despite both types of Dirac cones derive from the Zr d-orbitals, the stronger hybridization with the B atoms exhibited by the cone on the $M-\Gamma$ line makes it very sensitive to relative distance variations. External tensile strain expands the lattice and the Zr atom adjusts its vertical distance to the honeycomb lattice, and is able to vanish the Dirac point. Figure \ref{fig6} shows the evolution of the ZrB$_2$ electronic states with SOC under increasing tensile strain up to 2\%. The vertical distance to the center of the hexagon is progressively reduced from the original 1.48\AA\ down to 1.44, 1.41, 1.37, and 1.33\AA\ for an enlargement of the hexagonal unit vectors from 1\% to 2\%, 3\% , and 4\% respectively. Although the hexagons expand, providing more room to accommodate the Zr atom, the Zr-B bond length is barely modified. The main effect leading to the Dirac cone vanishing can be ascribed to the difference in the p$_z$ - d orbital hybridization as a result of the bonding angle modification. These structural changes, that leave the triangular geometry of the Zr atom intact, preserve the Zr-derived Dirac cone in the $\Gamma-K$ line. Although the cone is created with the participation of the B atoms that turn over the slope of the d$_{xz}$ derived band, it is less sensitive to hybridization effects. With no other metallic band to compensate the hole pocket, the Dirac point shifts to exactly the Fermi energy. Similarly, the cones at the $K$ point are separated by a slightly larger gap.

\begin{figure}[htp]
 \centering
 \includegraphics[width=0.5 \textwidth]{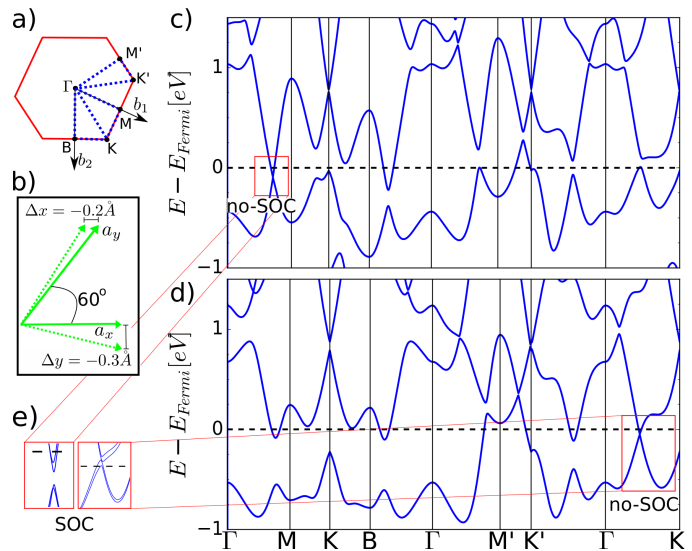}
 \caption{Dirac cones of monolayered ZrB$_2$ vanish selectively depending on the applied shear stress direction. a) shows the BZ obtained modifying the unit cell vectors as indicated in b). c) displays the non-relativistic band diagram obtained applying a force along the x-axis that modifies the $a_y$ vector x-component by -0.2 \AA. Similarly, d) is obtained modifying the $a_x$ vector y-component by -0.3 \AA.  (e) shows the modification that spin-orbit coupling introduces in the description of selected Dirac cones of the nonrelativistic band diagrams.}
 \label{fig7}
\end{figure}

A distinctive feature of graphene low-energy spectrum is that the Dirac cones at $K$ and $K$' points are protected by symmetry and from any perturbation that does not violate parity and time reversal symmetries. The metallic character of graphene electronic structure is thus insensitive to deformations such as external stress. On the contrary, the hexagonal network of ZrB$_2$ is highly sensitive to shear stress which reduces the symmetry of the unit cell and causes the vanishing of either the $\Gamma-K$ or $M-\Gamma$ Dirac points, depending on the direction of the external force. Figure \ref{fig7}a displays the shape of the BZ after symmetry reduction. The modification of the unit cell is realized by modifying one unit cell vector component at a time, as shown in Figure \ref{fig7}b, where the angle between in-plane unit cell vectors is increased. Figure \ref{fig7}c shows the electronic band diagram of ZrB$_2$ obtained upon applying an external force acting parallel to the x-axis that subtract 0.2 \AA\ to the x-component of the $a_y$ unit cell vector. Dirac cones in the $\Gamma$-M line split due to the SOC (see inset) and are preserved despite the symmetry reduction, while the Dirac point in the $\Gamma-M'$ line vanishes. Band degeneracy at the $K$ and $K'$ points is also removed. Similarly, decreasing by 0.3 \AA\ the y-component of the $a_x$ vector, a force acting along the y-axis is mimicked. As a result, only the Dirac cone in the $\Gamma-K$ line remains unaffected. 

\section{Conclusions}
The existence of twelve inequivalent Dirac cones in the low spectrum diagram of a hexagonal lattice can be realized, without resorting to Rashba SOC or twisted bilayer of triangular lattices\cite{PhysRevB.93.035401}. Owing to the formation of monolayered ZrB$_2$ Dirac cones in the lines joining high symmetry points, the cones are not shared with any neighbouring BZ. The twelve Dirac cones are divided in six groups of two non-symmetry-related cones with a slight electron-pocket compensation. First-principles calculations allowed to unveil the separate contribution of each d-orbital of the transition metal atom, and their hybridization with the B atomic orbitals. A color-scheme enabled to assign the formation of the cones to the Zr d$_{xy}$, d$_{xz}$, and d$_{yz}$ orbitals with a contribution from B p$_z$ orbitals as a result of atomic orbital hybridization.

Dirac cones are modified under external physical stimuli differently as a result of the unidentical participation of the B p$_z$ orbitals in the formation of the cones. Owing to the strong Zr-B atoms hybridization, small perturbations in the bonding angle lead to one cone vanishing. Ignoring SOC, internal symmetry breaking leads to a selective Dirac cone disruption at Fermi level depending on the direction of applied distortion, showing that not all cones are protected by symmetry. SOC induces complete Dirac point vanishing depending on the direction of the external force. These findings suggest that the interplay between d and p orbitals of different atoms, together with the relativistic effect of SOC, creates an environment that offers new avenues to electronics, and may provide a platform that outdoes the capabilities of other 2D materials with Dirac states.

It is worth mentioning that one key issue of monolayered ZrB$_2$ with a view to its implementation in electronic applications is the robustness of the electronic properties against the disorder introduced by intrinsic or extrinsic defects, such as atomic vacancies and external functional groups. Both short and long range disorder are known to limit the charge transport ability of 2D compounds\cite{ROCHE20121404} and, therefore, a systematic analysis of the possible role of defects in an energy-dependent charge mobility framework is proposed for future studies. From our results, where minute network deformations were shown to lead inevitably to drastic changes, the electronic properties of monolayered ZrB$_2$ can be anticipated as extremely sensitive to network changes. Thus, sensor applications by monitoring electrical changes in the nano-structure upon interaction with foreign species could be expected as a potential technological use of the material. Advanced approaches accounting for the strongly correlated Zr electrons must be considered to study the formation of localized magnetic moments upon composition changes. 

\section{Acknowledgments}
This work was supported by the U.S. Department of Energy, Office of Science, Basic Energy Sciences, Materials Science and Engineering Division. I acknowledge the computing resources provided on Blues and Bebop, the high-performance computing clusters operated by the Laboratory Computing Resource Center at Argonne National Laboratory. I thank P.B. Littlewood for fruitful discussions and support. 

\bibliography{biblio}{1}  
\end{document}